\documentclass[12pt]{elsarticle}
\usepackage{lineno}
\usepackage{slashed}
\usepackage{epsfig,graphics}
\usepackage{hyperref}
\usepackage{bbm}
\usepackage{mathtools}
\usepackage{amssymb}
\usepackage{mathrsfs}
\usepackage{amsmath}
\usepackage{color}
\newcommand{\panda}{$\overline{\textrm{P}}\textrm{ANDA}$\! }

\newcommand{\pandaPunkt}{$\overline{\textrm{P}}\textrm{ANDA} .$\,}
\newcommand{\pandaKomma}{$\overline{\textrm{P}}\textrm{ANDA} ,$\, }
\begin{document}
\begin{frontmatter}
\title{Resonances in QCD}

\author[GSI,TUD]{Matthias F.\ M.\ Lutz}
\address[GSI]{GSI Helmholtzzentrum f\"ur Schwerionenforschung GmbH, D-64291 Darmstadt, Germany}
\address[TUD]{Technische Universit\"at Darmstadt, D-64289 Darmstadt, Germany}

\author[giessen]{Jens S\"oren Lange}
\address[giessen]{II.\ Physikalisches Institut, Justus-Liebig-Universit\"at Giessen, \\D-35392 Giessen, Germany}

\author[JLAB]{Michael Pennington }
\address[JLAB]{Thomas Jefferson National Accelerator Facility, Newport News, VA 23606, USA}

\author{\\and}

\author[ferrara]{\\Diego Bettoni}
\address[ferrara]{Istituto Nazionale di Fisica Nucleare, Sezione di Ferrara, 44122 Ferrara, Italy}

\author[TUM]{Nora Brambilla}
\address[TUM]{Physik Department, Technische Universit\"at M\"unchen, D-85747 Garching, Germany}

\author[florida]{Volker Crede}
\address[florida]{Department of Physics, Florida State University, Tallahassee, FL 32306, USA}

\author[novosibirsk1,novosibirsk2]{Simon Eidelman}
\address[novosibirsk1]{Novosibirsk State University, Novosibirsk 630090, Russia}
\address[novosibirsk2]{Budker Istitute of Nuclear Physics SB RAS, Novosibirsk 630090, Russia}

\author[juelich]{Albrecht Gillitzer}

\author[mainz]{Wolfgang Gradl}
\address[mainz]{Institut f\"ur Kernphysik, Johannes Gutenberg-Universit\"at Mainz, \\ D-55128 Mainz, Germany}

\author[graz]{Christian B. Lang}
\address[graz]{Institut f\"ur Physik, Universit\"at Graz, A-8010 Graz, Austria}

\author[giessen]{Volker Metag}

\author[osaka]{Takashi Nakano}
\address[osaka]{Research Center for Nuclear Physics, Osaka University, Osaka 567-0047, Japan}

\author[ific]{Juan Nieves}
\address[ific]{Instituto de F\'isica Corpuscular (IFIC), Centro Mixto CSIC, Universidad de Valencia, E-46071 Valencia, Spain}

\author[heidelberg]{Sebastian Neubert}
\address[heidelberg]{Physikalisches Institut, Universit\"at Heidelberg, D-69120 Heidelberg, Germany}

\author[tit]{Makoto Oka}
\address[tit]{Department of Physics, Tokyo Institute of Technology, Tokyo 152-8551, Japan}

\author[korea]{\\Steve L. Olsen}
\address[korea]{Center for Underground Physics, Institute for Basic Science, \\Daejeon 305-811, South Korea}

\author[glasgow]{Marco Pappagallo}
\address[glasgow]{School of Physics and Astronomy, University of Glasgow, \\Glasgow G12 8QQ, United Kingdom}

\author[TUM]{Stephan Paul}

\author[bochum]{Marc Peliz\"aus}
\address[bochum]{Institut f\"ur Experimentalphysik I, Ruhr-Universit\"at Bochum, \\D-44801 Bochum, Germany}

\author[rom,JLAB]{Alessandro Pilloni}
\address[rom]{Dipartimento di Fisica, Universit\`a di Roma La Sapienza, 00185 Roma, Italy}

\author[juelich]{Elisabetta Prencipe}
\address[juelich]{Institut f\"ur Kernphysik, Forschungszentrum J\"ulich GmbH, D-52425 J\"ulich, Germany}

\author[juelich]{Jim Ritman}

\author[dublin]{Sinead Ryan}
\address[dublin]{School of Mathematics, Trinity College, Dublin 2, Ireland}

\author[bonn]{Ulrike Thoma}
\address[bonn]{Helmholtz-Institut f\"ur Strahlen- und Kernphysik, \\Rheinische Friedrich-Wilhelms-Universit\"at Bonn, D-53115 Bonn, Germany}

\author[heidelberg]{Ulrich Uwer}

\author[ECT,TUM]{Wolfram Weise}
\address[ECT]{ECT*, Villa Tambosi, 38123 Villazzano (Trento), Italy}

\begin{abstract}
We report on the EMMI Rapid Reaction Task Force meeting '{\it Resonances in QCD}', which took place at GSI October 12-14, 2015 (Fig.~1).
A group of 26 people met to discuss the physics of resonances in QCD. The aim of the meeting 
was defined by the following three key questions:
\begin{itemize}
\item What is needed to understand the physics of resonances in QCD?
\item Where does QCD lead us to expect resonances with exotic quantum numbers?
\item What experimental efforts are required to arrive at a coherent picture?
\end{itemize}
For light mesons and baryons only those with {\it up}, {\it down} and {\it strange} quark content were considered. For heavy-light and heavy-heavy meson systems, those with {\it charm} quarks 
were the focus.

 This document summarizes  the discussions by the participants, which in turn led to the coherent conclusions we present here.
\end{abstract}
\end{frontmatter}


\section{Introduction}

After many decades of theoretical and experimental work, low energy QCD remains the most 
challenging frontier in the physics of the Standard Model. 
The Lagrangian of QCD is simple, yet from its strong
coupling dynamics emerge  all the phenomena of the nuclear and hadron world.
How colour confinement, chiral symmetry breaking, the properties of the hadron spectrum, and inter-nucleon forces
result from the strong dynamics encoded in QCD defines the low energy frontier.

A most striking phenomenon of QCD is the formation of the constituents of ordinary 
baryonic matter, the proton and the neutron, out of massless gluons and almost massless 
quarks. To gain insight into this complex dynamics the baryonic excitation spectrum has been 
studied for many decades. 
Significant progress has been achieved with the recent availability of new polarization data, with polarized beams and/or polarized targets. 
While there is an increasingly rich collection of empirical data on photon-nucleon reactions, 
the world data on the key pion-nucleon and kaon-nucleon reactions are more limited. Each provides complementary
information on the baryon resonance spectrum of QCD.  Even less exploited are 
antiproton-nucleon, proton-proton and $e^+e^-$ reactions for the study of baryon resonances.  

Unfortunately the available empirical data are not matched by a deep theoretical understanding 
of such reactions from first principles QCD. Though QCD lattice simulations have progressed 
significantly over the last decade, reducing statistical uncertainties and employing robust techniques for 
spin identification  \cite{Edwards:2011jj}, a calculation of the physical excited baryon spectrum is still 
a tough challenge with present computing power.

The key question is what are the relevant degrees of 
freedom for the resonance physics of QCD. Are the so-called constituent quarks, quasi-particles, whose mass is a 
consequence of the spontaneously broken chiral symmetry of QCD, the most efficient way to describe reaction amplitudes 
and the excitation spectrum of QCD with light quarks? Though the simple quark model has had many successes, it is 
more and more challenged by the steadily increasing empirical data set. Examples of unconventional states are 
the light scalar states, the $\Lambda(1405)$, the $N^*(1440)$, the $D_{s0}^*(2317)^+$, the $XYZ$ states and more.
To what extent are diquark correlations, gluonic modes or hadronic degrees of freedom important in this physics? An extreme point of 
view would be the hadrogenesis conjecture, 
which attempts to describe the spectrum in terms of a selected set of hadronic degrees of freedom, very much like nuclear structure 
physics that is very successfully described in terms of just the proton and neutron as the effective degrees of freedom. 

For long it had been thought that states with hidden heavy flavour were simpler. Both charmonium and bottomonium states could be described 
as $\bar q \,q$ systems (with $q = c$ or $b$) bound 
by a confining potential. Effective Field Theories have allowed us to obtain such potentials,
static and with relativistic corrections, directly from QCD and to calculate
them at high order in the perturbative expansion for small quark-antiquark
distance or on the lattice in the large distance region \cite{Brambilla:2004jw}. In this way the properties of the lowest
quarkonia resonances are understood in terms of QCD
\cite{ Brambilla:2010cs}.
This picture however is valid
only away from the strong decay threshold, where new
degrees of freedom enter and a different picture arises.
In fact, new data on charmonium-like systems have for the first time 
unambiguously  established the existence of exotic resonant states that 
involve a minimal configuration of four quarks. The $\bar q q$ potential model is too simple:
more degrees of freedom must be effective. This should have important implications for the resonance physics of 
{\it up}, {\it down} and {\it strange} quarks too. It was the purpose of the EMMI task force to scrutinize such consequences for selected 
sectors of QCD. The main mission was to work on the following three questions

\begin{itemize}
\item What is needed  to understand the physics of resonances in QCD?
\item Where does QCD lead us to expect resonances with exotic quantum numbers?
\item What experimental efforts are required to arrive at a coherent picture?
\end{itemize}

A diversity of experimental programmes on resonance physics in QCD is  presently going on or planned for 
the next few years. Moreover, future work using lattice QCD simulations requires detailed planning to meet the increasing demand 
for such calculations both in terms of computer power and manpower resources. 

In order to focus the task force meeting, discussions 
were restricted to four specific areas of QCD. Light mesons and baryons with {\it up}, {\it down} and {\it strange} quark content were 
considered. In addition heavy-light and heavy-heavy meson systems with {\it charm} quarks were discussed. 
For all four sectors an attempt was made to identify the most promising key experiments which should lead 
to progress in the understanding of resonances in QCD. 
Members of different collaborations  were invited, in particular of Belle, BESIII, COMPASS, GlueX, JPARC, LHCb and PANDA
being key players in this field.
This was  complemented by  a group of leading theoreticians with a balance of 
phenomenology and lattice experts.
In this short report the referencing can only be selective. There is no claim of completeness and we implicitly assume 
the relevance of the references embodied in the works we cite.

\section{Day 1: Light mesons}
\vskip0.3cm
\noindent

We are far from having a profound understanding of the meson spectrum composed of {\it up}, {\it down} and {\it strange} quarks. Experimental 
data indicate a possible proliferation of states compared to the simple quark model picture. In the low mass region, 
chiral symmetry appears to describe the first scalar and axial-vector states \cite{GomezNicola:2001as,Lutz:2003fm} quite successfully. However, how 
to link this to
the empirical spectrum at higher masses and unravel the driving degrees of freedom of QCD remains a challenge. Here lattice QCD simulations are
starting to have critical impact: scattering phase shifts can be extracted from 
energy levels of finite box QCD at least where only a few hadronic channels intrude. The issues considered cover: are there glueballs \cite{Sanchis-Alepuz:2015hma}, hybrid, 
4-quark \cite{Eichmann:2015zwa} or molecular states \cite{GomezNicola:2001as,Lutz:2003fm}? How far do the Regge trajectories 
extend \cite{Fischer:2014xha} with increasing mass and spin? 

Light mesons are promising systems to search for gluonic degrees of freedom. 
The gluon self-interaction is a defining element of QCD. It allows bound states of gluons, even in a world without quarks. From the earliest models and calculations the ground state ``pure'' glueball is expected around 
a mass of 1.5 GeV. Such a mass is readily reached experimentally in several production mechanisms which are believed to provide gluon-rich environments, such as $J/\psi$ decays, central 
production in proton-proton collisions, or proton-anti-proton annihilation. However, the experimental evidence so far is 
inconclusive, or at least controversially discussed. The inevitable strong mixing with conventional states of identical quantum 
numbers makes the unambiguous identification of glueballs a major challenge.  
A systematic study of the spectrum and the properties of mesons up to higher masses is needed to establish if there are 
indeed any supernumerary states indicative of glueballs.
 
The calculation of the glueball spectrum with dynamical quarks is a challenge requiring very high statistics as well as disentangling 
the allowed meson-glueball mixing. For now, the quenched calculation~\cite{Morningstar:1999rf} remains 
a benchmark. In unquenched calculations it is difficult to disentangle the consequences of
meson and glueball operators. 
As a consequence, clear predictions for decay rates, and 
mixing with conventional scalars, are missing at present, pending further study.


\vskip0.5cm
\noindent{Experimental studies}
\vskip0.3cm

Diffractive production as well as the decays of $J/\psi$, $\psi(2S)$, $D$ and $D_s$, or even $\eta_c$, offer a great laboratory 
to study scalar/tensor isoscalar mesons and isovectors too. The production of glueballs in $J/\psi $ radiative decays at BESIII seems 
small despite the expectation that this is a gluon-rich system. Does  $p\,\bar p$  annihilation provide a better glue-rich environment?
The observation of both the scalar $f_0(1500)$ and the  pseudoscalar $\eta(1405)$ glueball candidates at LEAR with
 Crystal Barrel and OBELIX suggest that $p\,\bar p$ is indeed a good reaction for glueball searches.

Is it easier to detect hybrid mesons than glueballs? 
Since many decay channels are kinematically allowed, both may have large widths. Thus, a detailed Partial Wave Analysis (PWA) is the only reliable approach
to extracting their signals. 
While the light glueballs all have conventional quantum numbers, many hybrid states may have quantum numbers not allowed in the quark model. These would be more readly identified in
so called  ``exotic'' waves in the PWA.
The upcoming programme at Jefferson Lab has been strongly motivated by lattice predictions~\cite{Dudek:2010wm} of a spectrum of 
hybrid states  in the 1.6-2.4 GeV region.
The GlueX experiment will search in photoproduction, while CLAS12 will use electroproduction.
The process of $p\,\bar p$ annihilation has no final state baryons which
complicate the analyses. Indeed \panda complements and extends other experimental programmes and has
the ability to confirm
results that might otherwise be controversial. Moreover, the facility to access higher-mass systems, where most of the hybrids are expected to be,
is unique. Even though the rates at \panda will be high ($\geq$10$^7$ candidate events per day),  
the very significant background of light meson production with identical kinematical signatures will 
require sophisticated analysis techniques, such as a multi-variate classifier,  to improve the signal-to-background ratio. 
In addition \panda will have the capability to search for states with gluonic degrees of freedom (hybrids and glueballs) in the 
charmonium energy region, where the overlap with conventional states will be much smaller, 
making the background situation much more favourable.

In any experimental study it is important to reconstruct both charged and neutral final states. 
An example is the $a_1(1420)$, with $J^{PC} = 1^{++}$, recently observed by COMPASS in its charged state \cite{Adolph:2015pws}.  
The $a_1$(1420) is a rare signal, representing only 0.25\% of the total intensity in that mass region.
The observation of this state relies on sophisticated Amplitude Analysis techniques, including up to 88 partial waves, and making use of the full 5-dimensional distribution of the three-body final state, $\pi^-\pi^-\pi^+$. The identification of its neutral partner using the $\pi^0 f_0(980)$ mode would help in establishing this as an isotriplet resonance.

The COMPASS results  have highlighted how all experiments must prepare for the robust extraction of small signals if they want to definitively identify 
new states such as hybrids and glueballs. This is an area of fertile cooperation between experimentalists  and theorists necessary to ensure the required methodologies are reliable
and firmly routed in current understanding of reaction theory and the analytic $S$-matrix. 
This is essential if states like the $a_1(1420)$ are to be confirmed as members of the hadron spectrum and not complex kinematical effects.

\section{Day 1: Open-charm meson systems}
\vskip0.3cm
\noindent

Meson resonances composed of a heavy quark and a light antiquark are further promising
systems for unraveling the dynamical degrees of freedom underlying how  structure is formed in QCD.
From a theoretical point of view such systems are unique, since here
chiral symmetry for the light quarks and heavy-quark symmetry 
are both manifest. These constrain the nature of the  internal interactions significantly. Indeed, this is the only system where
coupled-channel dynamics based on the leading order chiral Lagrangian predicts the existence
of an $SU(3)$ flavour multiplet of resonance states with exotic quantum numbers \cite{Kolomeitsev:2003ac}. Clear measurable
signals in specific final states are foreseen. So far these predictions have neither been disproven nor
confirmed by experiment. A detailed study of the open-charm meson spectrum is expected to provide
important clues for a deeper understanding of resonances in QCD.

LHCb has recently illustrated the power of using exclusive decays of $B$-mesons for the identification of excited charmed mesons.
Amplitude Analyses based on an assumed isobar model with excellent signal-to-background ratio  have been performed.
Exploring masses up to almost 3 GeV, LHCb has been able to identify a $D_{sJ}$ with spin as high as $J=3$.
So far, BaBar und Belle have not been able to find D mesons heavier than 3 GeV either.
While current studies are dominated by charged particle final states, decays into neutral mesons are critical to understanding the 
dynamics of these states.
Belle II will explore how to complement current results by data with neutral particles. 
Final states involving a $\eta \,D$ or $\eta \,D^* $ are important
channels to discover a possible exotic flavour sextet. Such states may decouple from the $\pi \,D$ or $\pi\, D^*$ channels,
but are expected to have a large branching fraction into channels with $\eta$ mesons, making the ability to detect the 
decay $\eta \to\gamma\gamma$ crucial.


Lattice QCD predictions for the low-mass spectrum are underway for heavy-light mesons. For instance,
first results on the low energy scattering phase shift for $D \,K$ were presented at the Task Force meeting \cite{Lang:2014yfa}.
It should also be noted that exploratory lattice QCD studies suggest so far unseen $D_s$ meson hybrid states at masses
$\geq$~3.4~GeV \cite{dsj_hybrids}. This will be an important search at current and future experiments.

A stringent test of our understanding of the workings of QCD is posed by narrow states. For instance, for the well established 
$D_{s0}^*(2317)^+$, experiment currently provides an upper
bound of a few MeV for its total width. Any hadronic decay is forbidden by isospin symmetry. Depending on different dynamical
assumptions, decay widths from 10 keV to more than 100 keV are predicted \cite{Lutz:2007sk,Cleven:2014oka}. Here a careful 
discrimination of the various theoretical
calculations is required. Over the last decade an amazingly consistent hadronic decay pattern was worked out relying on the
flavour $SU(3)$ chiral Lagrangian, including the charmed meson ground states with $J^P=0^-, 1^-$. 
The width of the $D_{s0}^*(2317)^+$ is then predicted to be around
140 keV by theory. The mechanism behind this is a mixing of the conventional flavour triplet with the exotic flavour sextet
that arises in the presence of isospin violating effects. Since in the flavour sextet channel the chiral Lagrangian predicts
significant attraction, the matrix elements in the sextet channels are large. This explains the unexpectedly large
predicted width of the $D_{s0}^*(2317)^+$ of $\geq$100~keV. 
Moreover, this chiral Lagrangian approach appears consistent
with recent lattice QCD simulations in this sector, which provide constraints on some low energy phase shifts.
High precision measurements of the width are needed to scrutinize this picture. Remarkably, \panda is expected to go down to about 100 keV by means of a 
threshold scan in $p\,\bar p \rightarrow D_s \bar D_{s0}$(2317). No other experiment can be that precise. 
In order to perform this measurement, analysis tools need to be developed to improve  the signal-to-background ratio, 
which for this channel is of the order of $1/10^6$. 

\section{Day 2: Light baryon systems}
\vskip0.3cm
\noindent

Our ultimate aim is to clarify what are the driving degrees of freedom for baryon resonances. That these are not always just 
three constituent quarks has been highlighted by the fact that
some very well established baryon states have for long proved an enigma. They are difficult to reconcile with a simple quark model 
picture, or even lattice calculations albeit with
heavy pions. For instance, the  
$\Lambda(1405) $ with 
$J^P=1/2^-$ and the Roper $N^*$ with 
$J^P=1/2^+$ do not fit 
the conventional picture of three quarks. Should we expect doubly strange baryons with similarities to these? 
Indeed a decade ago a series of studies showed that the lowest baryon states 
with
$J^P =1/2^-$ and $3/2^-$
are naturally predicted by the chiral Lagrangian formulated for three 
light flavours \cite{GarciaRecio:2003ks,Kolomeitsev:2003kt}. Therein 
the  Goldstone bosons couple to the baryon octet and decuplet ground state fields with $J^P = {1}/{2}^+$ and $J^P = {3}/{2}^+$, respectively. If the leading-order chiral interaction is used as input in a coupled-channel 
calculation, states like the $\Lambda(1405) $ will be dynamically generated. This mechanism is analogous to the one which 
describes the lightest scalar and axial-vector mesons (with or without {\it charm}) from the chiral Lagrangian. Such studies have shown  
that the $\Lambda(1405) $ is not necessarily an exceptional state, as was argued for long in the literature, rather it may 
be more typical among excited states. However, this prediction awaits experimental confirmation in other strangeness sectors. It is an unsolved puzzle 
how this consequence of QCD squares with the quark-model picture of baryon resonances. The role of the approximate chiral flavour $SU(3)$ 
symmetry remains largely a mystery, even though it has been illustrated that this symmetry seems 
to be a feature of at least part of the excitation spectrum. Is a similar dynamical mechanism applicable to high-mass 
baryon resonances? What is the role of the light vector mesons, which are suggested to be of crucial relevance by the hadrogenesis 
conjecture? 

Another approach to the excited baryon spectrum relies on properties of QCD in the limit of a large number of colours ($N_c$).
In this limit it is argued that the spectrum exhibits simpler symmetry properties \cite{Schat:2001xr,Goity:2002pu}. Back at the physical point 
with $N_c=3$ one may expect an approximate $O(3)\otimes SU(6)$ symmetry, which allows the classification of the spectrum in multiplets. The 
current knowledge of the $N^*$ and $\Delta^*$ spectrum seems to leave out a 20-plet (a flavor octet with 
$J^P=1/2^+$ and  
a flavor singlet with 
$J^P=3/2^+$). What does this imply for the strange baryon resonances? We note that large-$N_c$ QCD is currently 
also used to systematically constrain counter terms in the chiral Lagrangian \cite{Dashen:1993ac,Lutz:2010se}. That renders coupled-channel 
computations based on the chiral Lagrangian highly predictive. Such constraints are  also important in studies of the quark-mass dependence 
of the baryon ground-state masses \cite{Lutz:2014oxa}.


Pioneering lattice studies for the highly excited baryon states in finite volume QCD appear to agree qualitatively with an excitation 
spectrum expected from the constituent quark model, or more generally from the multiplet pattern predicted by 
large-$N_c$ QCD \cite{Edwards:2011jj}.
However, this is to be taken with a grain of salt. These studies by the {\it Hadron Spectrum Collaboration} from a few years ago~\cite{Edwards:2011jj} rely on using only 3-quark operators 
that were constructed with light quarks that are unphysically heavy. 
 Meson-baryon interpolating operators have to be included, particularly as calculations proceed to lighter quark masses. 
 Then the excited baryon states will decay and
 using the L\"uscher connection of the finite to infinite volume $S$-matrix, scattering phase shifts 
should be extracted. The spectrum deduced from such coupled-channel studies may be 
significantly modified
compared to the exploratory results~\cite{Verduci:2014csa}. Within the next few years one may
expect further results on phase shifts. 
Branching ratios may then become available for some resonant states. Nevertheless, phase shifts for light baryons much above 
2 GeV will remain difficult to obtain from lattice QCD because of the many open decay channels.


\begin{figure}[t]
\center{\includegraphics[keepaspectratio,width=1.0\textwidth]{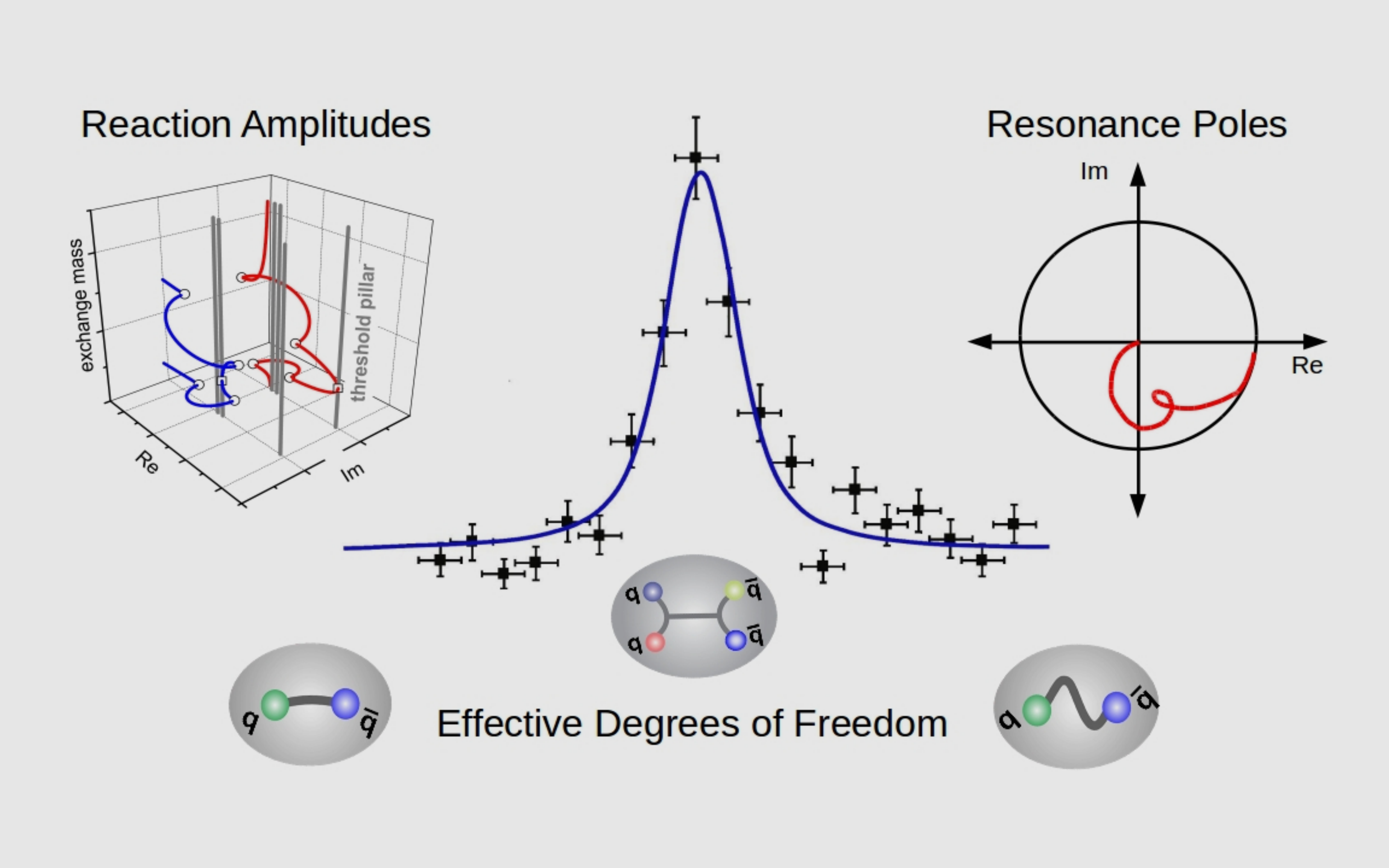}%
}  %
\caption{\label{fig:1} Figure taken from the poster of the EMMI Rapid Reaction Task Force 'Resonances in QCD'. See  
indico.gsi.de/conferenceDisplay.py?confId=4058. }
\end{figure}

\vskip0.5cm
\noindent{Experimental studies}
\vskip0.3cm

Using photoproduction for baryon spectroscopy relies on well understood production amplitudes from electromagnetic interactions.
In the last decades ELSA, MAMI, JLab and LEPS have collected a huge data set on resonance production, complemented 
by measurements with a polarized target and/or a polarized beam \cite{Crede:2013sze}. The polarization data have made a dramatic 
impact on state-of-the-art Partial Wave Analyses. They have uncovered new nucleon and $\Delta$ resonances that were not previously revealed in 
the analysis of experiments with pion beams \cite{Anisovich:2012,Sokhoyan:2015eja,Hartmann:2015kpa}. 
However, the ability to photoproduce and then identify excited  baryons much above 2~GeV may be difficult.
A complementary source for nucleon resonances is the $J/\psi$ factory of the BESIII experiment, which is already used to study 
excited $N^*$'s from dedicated partial-wave analyses \cite{Ablikim:2012}.

Baryon resonances with strangeness are significantly more difficult to study and the available data set is rather limited. This is particularly so 
for doubly- and triply-strange systems. So far empirical information stems mainly from reactions with either photon, pion or kaon beams, 
where a certain number of kaons have to be tagged in order to access baryonic systems with negative strangeness. For the few 
established doubly-strange states  little is known about their $J^{P}$ assignments. In a simple quark model picture, the strange states will 
systematically fit into the appropriate multiplets as those of the $u, d$ sectors. However, it could be 
that the dynamics of excited baryons differs from that of the lower lying states. Their pattern of decays may be systematically different. 
Parity doublets may appear in some sectors with increasing mass.  It is therefore of crucial importance to 
collect a significant data base to provide decisive physics clues from studies of baryons with strangeness. 
Indeed for the crypto-exotic $N^*$ ($P_c$) recently discovered around 4.4 GeV by LHCb \cite{Aaij:2015tga}, 
the $\Lambda^*$ resonance states contribute to a significant kinematical reflection in the region of the putative pentaquarks producing a strong interference pattern.
 A reliable extraction of the exotic signal requires a detailed 
understanding of the $\Lambda^*$ spectrum.

JLab has presented plans for a dedicated search for double strange baryons with GlueX and CLAS12. A reliable estimate of the 
production rates is difficult. 
In the existing data the number of peaks due to $\Xi$-resonances seems to be quite limited. Larger rates are foreseen with \pandaKomma which 
is expected to support a dedicated programme to search for doubly-strange baryons \cite{Lutz:2009ff}, with an accompanying partial wave analysis. 
Moreover, the rate estimates for baryons with three strange quarks are promising and warrant dedicated studies. 

LHCb and Belle have presented interesting data sets on $S=-1, -2, -3$ baryon ground states. These data sets should be extended to search 
for excited states. Hyperon spectroscopy is not the focus of either LHCb or Belle. 
Thus, it is difficult to estimate the discovery potential. 
While the strange baryon ground states can be cleanly identified owing to 
their long weak-decay lengths, excited states decay by single or multi-pion emission and therefore are more difficult to 
study. The identification of resonance states requires the use of PWA tools. This will be a challenge at LHCb owing to a) the combinatorial 
background in the large multiplicity environment and b) the absence of a clean production process. 
Most promising are exclusive $b$-hadron decays that allow a PWA based on a well defined 
initial state.  Belle II multiplicities are much lower than those of LHCb. Hyperons are produced in part 
from the continuum and may stem from exclusive reactions (in addition to $B$-decays). In particular pair-production of baryons may 
be a good laboratory for hyperon studies.  
COMPASS  has observed strange baryons in exclusive $p\, p$ scattering with $K\,\bar K$ pairs in the final state. The  $S=-1$ baryons 
could be addressed this way. 

Complex analysis tools are being developed for LHCb, Belle and COMPASS, but must be further refined to possibly 
identify decays of excited hyperons.

\section{Day 3: Charmonium-like systems}
\vskip0.3cm
\noindent
It is the charmonium-like sector that has excited the interest of the hadron and nuclear community most widely.
Here thanks to recent progress in both effective field theories
and lattice calculations,  the properties of the lowest resonances have been understood
  directly from QCD with  great precision and in terms of dominant quark-antiquark
degrees of freedom.
Yet close to or just above the strong decay threshold, there has been an
explosion of new discoveries by BESIII, BELLE, Babar (some confirmed by Fermilab)
and now the LHC experiments. Indeed, these have provided strong evidence that more complex 
configurations allowed by QCD contribute to the observed spectrum with new resonances called $X Y Z$ states. 
Many phenomenological models with additional degrees of freedom have been
developed to describe such resonances.
Possibilities are (a) tetraquarks, 
with e.g. diquark anti-diquark forces contributing to the binding, (b) weakly bound molecules of open-charm mesons, (c) 
excitations of light quarks orbiting a heavy quark pair in the centre (often called ''hadro-charmonium''), or (d) hybrids, 
with excitations of the gluon flux tube between the two heavy quarks. These assumptions provide different excitation patterns, 
which can be compared directly to the experimentally observed patterns \cite{Brambilla:2014jmp}. First exploratory 
lattice results are available \cite{Liu:2012jr,Prelovsek:2015fra}. Recently a QCD derived effective field theory description for hybrids has
been formulated \cite{Berwein:2015vca}.

One of the most important tasks is to map out the pattern 
of $XYZ$ states, in particular their yet unobserved (spin) partners. 
For instance, in the hadrocharmonium model \cite{hadrocharmonium}, one may expect the $Y(4260)$ to be 
related to the $J$/$\psi f_0(980)$ channel.  Moreover, the $D_1(2420) \bar D$ threshold is quite close and may also play 
an important dynamical role \cite{Wu:2013onz}. Depending on the assumed internal configurations (a)--(d) above, the pattern of partner states is different. 
For instance, in the $P$-wave tetraquark model, the  1$^{--}$ $Y(4260)$ would be degenerate with a $3^{--}$ state \cite{Cleven:2014oka}, which would be 
accessible at \pandaPunkt  Angular momentum barrier effects may make it difficult to observe such states in $B$-meson decays. Consequently, the complete pattern of partners may have to await the running of \pandaPunkt 

Some of the charged $Z$ states have been observed in $B$ decays, while others are seen  in $e^+e^-$ collisions,
in particular in $Y(4260)$ decays. However, no $Z$ state has been observed in both production mechanisms. This suggests
 that other production processes, such as $p \,\bar p $ collisions, may be  required to understand why. 
Assuming an integrated luminosity of 0.5 pb$^{\rm -1}$/day, which corresponds to the low luminosity mode 
with 10$^{\rm 31}$ cm$^{-2}$ s$^{-1}$ for the start of \panda data taking, significant numbers of $X(3872)$, $Y(4260)$ and $Z^+(3900)$  
are expected already in year one. Cross-sections are in the order of nanobarn, compared to $e^+e^-$ collisions \cite{lange_charm13,prencipe_hadron15}.


One of the prime examples for an exotic, charmonium-like state remains the $X(3872)$. 
Even a decade after its discovery its nature remains unresolved \cite{Braaten:2014qka,Guo:2014xrk}.  
We identify three important tasks for upcoming experiments:
the very close vicinity of the $X(3872)$ to the $D^0\,\overline{D^{*0}}$ threshold is 
one of the most striking observations, possibly pointing to a large molecular component in its wave function. 
A significant signal of the $X(3872)$ has been observed in the decay 
$B^+\!\!\rightarrow \!K^+$X(3872) \cite{b2kdd_babar} \cite{b2kdd_belle}.
However, to our knowledge, there has not been a complete study 
of $B^+ \!\rightarrow \!K^+D^{*0} \,\overline{D^{*0}}$ or $B^+\!\rightarrow \!K^+D^{*+}{D^{*-}}$, 
due to limitations in the reconstruction of low momentum $\pi^\pm$ and $\pi^0$. 
The latter decays would be the primary search channels for a possible 2$^{++}$ partner of the $X(3872)$ 
at the $D^{*0}\,\overline {D^{*0}}$ threshold \cite{Albaladejo:2015dsa}. 
We expect both LHCb and Belle II to contribute to this search in the near future.
For the width of the $X(3872)$, an upper limit of 1.2~MeV is presently determined  \cite{pdg}.
If the Fock decomposition of the $X(3872)$ contains molecular components, its width may be significantly larger than the width of the $\psi$(2S) with $\Gamma(\psi(2S))$ $\simeq$ 300~keV.
Belle II may be able to reach
this value with a multidimensional, kinematically 
overconstrained fit. Indeed, if the $X(3872)$ is purely molecular, then there is a direct relation between its width and its binding energy~\cite{Polosa:2015}, giving a precise determination of its width added importance.

Rare decays of the $X(3872)$ are an opportunity for \panda.  The enormous statistics provided by $10^4$ $X(3872)$'s produced per day, even 
at the start of data taking, makes such studies feasible. 
Decays to light mesons should be OZI suppressed 
if the $X(3872)$ is a pure $c\,\bar c$ state. 
If the X(3872) contains tetraquark eigenstates in its Fock decomposition,
light quark rearrangement might enhance the branching fractions.
While  radiative transitions from and to the $X(3872)$ are suppressed 
by $\alpha_{em}\simeq 1/137 $, they nevertheless provide insight into the complete 
spectroscopic pattern of the state \cite{Aaij:2014ala}.
Colliding antiprotons on nuclei with \panda would allow the  $A$-dependence
of the production of $\psi'$ and $X(3872)$  near threshold
to be compared. This may, after appropriate theoretical study, provide a good way to expose an extended $D^*\bar D$
component of the $X(3872)$ state function.

\section{Summary and conclusions}

The task force was organized into public morning sessions and closed afternoon brainstorming discussion rounds. 
While in the morning the topic was introduced with overview talks, in the afternoon short contributions were 
prepared to help the discussions. The main guide for the afternoon discussions were the two questions

\begin{itemize}
\item What do we need and why?
\item How do we get there?
\end{itemize}

\noindent Though the discussions were naturally divide into theoretical and experimental issues, there was an important synergy between them. 


On the theory side there appears a  consensus.
Significant progress from lattice QCD simulations has been made. Very recently the community started to 
compute scattering phase shifts from the set of energy levels in a finite box. In order to apply L\"uscher's framework studies 
on several (and larger) lattice volumes
including coupled channels (see, e.g.,~\cite{Dudek:2014qha,Ryan:2015gda}) are required.
In  the alternative HAL QCD pseudo-potential approach~\cite{Sasaki:2015ifa} the HAL QCD 
collaboration is setting up configurations in boxes of spatial sizes up to 10 fm. The determination of the 
phase shifts will involve complicated coupled-channel dynamics. 
Here it is important to consider source fields 
that couple the quark-gluon dynamics to all important hadronic final states, as already done in the simpler $\pi\pi$ and $\pi K$ sectors, 
\cite{Dudek:2014qha}.  The HAL-QCD has attempted to compute and apply multiple-channel potentials to calculating the $S$-matrix for exotic 
charmonium-like mesons. A few examples were shown at the meeting 
of the striking effects of including such sources on the volume dependence of the energy levels. 
Further attention to explore channel coupling effects in lattice QCD calculations is necessary to bridge the \lq\lq finite-volume discrete states" 
on the lattice and the physical resonances made of a variety of open and closed channels. 
A close interaction of the lattice community with experts on hadronic final state interactions is already very fruitful. 
The challenge is to compute reaction amplitudes, extract pole positions on higher Riemann sheets and study the quark mass-dependence 
of the resonance properties.

In addition, to be able to deduce a physical picture from experimental or lattice data we need QCD motivated models. Here it is important to work 
out extreme cases that help discriminate the distinct patterns associated with a specific choice of degrees of freedom. 
Model calculations aiming at a description of the $XYZ$ spectrum should reproduce and predict patterns of excited 
states with their branching ratios rather than describe the properties of a few selected states. From such studies 
we want to learn what are the most effective coloured and/or colour neutral degrees of freedom in the resonance physics of QCD. 

Different sectors elucidate  different properties of QCD. In the charmonium sector the confinement of heavy quarks is probed. On the other 
hand, for a system composed out of {\it up}, {\it down} and {\it strange} quarks, the details of colour confinement are as yet poorly understood. 
The dominant QCD property here may be its spontaneously broken chiral symmetry. Quantitative insight can be gained from the 
limit of QCD where the number of colours approaches infinity. How such properties are reflected in hadronic final state interactions
is a crucial challenge the theory community needs to address vigorously.  

Heavy-light systems composed of a {\it charm} quark and a light quark ($u,d$ or $s$) are constrained not only by chiral symmetry, but also by the 
heavy-quark spin symmetry. Such mesonic systems are particularly exciting. It is here that chiral $SU(3)$ symmetry suggests 
a flavour sextet of states, that cannot occur in a quark-antiquark picture. This prediction can be probed with 
lattice simulations.


\vskip0.5cm
\noindent{Experimental needs}
\vskip0.3cm
In order to make progress in the understanding of these hadronic states,
spin-parity and full partial wave analyses are required, which require high statistics data samples, large angular acceptance and as
complete particle identification as possible.
Assuming purely experimental issues of statistical and systematic apparatus effects are being solved,
the quest is for complementary reaction processes in order to narrow down the interpretation of new states
being, or to be, observed. For instance,  production by $t$-channel exchange, which allows clean partial wave analyses, is characterized
by coherent overlap of various exchange processes, while production in $s$-channel reactions, or decay processes, are typically amenable to Dalitz plot analyses.
Both are required to
shed light on the interpretation of resonances observed (generic vs reaction specific resonance interpretation).
Similarly, continuum ($s$-channel) production in colliders must be complemented with studies of heavy flavour or heavy lepton decays --- processes with similar systematics.
 Such complementary approaches are mandatory for the determination of pole parameters, for which
an extrapolation in the complex energy plane could well be ambiguous in single experiments. It is therefore crucial that the various data sets are
published so as to allow combined fits, and that plans are made for  the tremendous efforts required from individual collaborations to be carried out jointly.



Currently COMPASS is enlarging the data base on light-meson spectroscopy \cite{Grube:2015fac,Adolph:2015tqa}.
The hunt for gluonic degrees of freedom in spectroscopy is ongoing. With GlueX a dedicated search for
light hybrid states has started. 
We have seen exciting and unexpected new results from LHCb using an 
unprecedented data sample with $\sim$$10^{13}$ events of directly produced open charm mesons.
So far no unambiguous exotic signals have been seen. In the hidden charm sector, LHCb has recently claimed novel pentaquark states. 
In all cases, it is important to extend such studies to
neutral channels. Here Belle II and \panda have experimental advantages and their  discovery potential for charmed hybrids 
will be particularly high for narrow states.

Recent data on polarization observables in photo- and electro-production experiments from ELSA, MAMI and JLab have had a huge impact on excited
nucleon and $\Delta$ studies. Such spin observables have led to more reliable Amplitude Analyses and have revealed new resonant states.
Here it has proved very successful to combine data from different experiments. 
High quality pion and kaon beam data are expected from J-PARC and will further
stabilize the current PWAs. Complementary information comes from $J/\psi $ decays
studied at BESIII, but presently this has not been included in more global  analyses. The strange baryon spectrum requires further experiments if a systematic
 detailed knowledge of strangeness $-1$ to $-3$ is to be acquired. There are plans at JLab for a dedicated search for double strange baryons with GlueX and CLAS12, and forthcoming studies at J-PARC.
Even larger production rates are estimated for \pandaKomma which should make possible the identification of
the high-mass strange partners of the newly observed nucleon and $\Delta$ resonances. This is essential for an understanding of the flavour structure of
the highly excited baryons.

In all the sectors,  it remains a challenge to sharpen the tools required to analyse the present and future data sets. Theoretical work is under way
to develop multi-channel analyses that go beyond a simple isobar treatment, and extend the $K$-matrix model of 2-body interactions to many-body final states. 
 Such multi-channel frameworks fulfill coupled-channel unitarity for more than $2\to 2$ reactions as well as 
respect the condition of micro-causality. The latter implies specific analytic properties of the partial-wave 
amplitudes \cite{Kennedy-Spearman:1962,Eden_Landshoff_Olive_Polkinghorne,Lutz:2015lca}. 
This provides a basis for distinguishing between what effects may be kinematical  and which are really due to states in the spectrum of QCD.
More widely, such methodologies are required to extract definitive information from data, whether from experiment or from lattice simulations.
Intense interaction between experiment and theory
has started with the NABIS, HaSpect and JPAC programmes.  


Numerous recent experimental observations of narrow charmonium-like states ($XYZ$ states) have
been reported from BESIII and Belle but also LHCb \cite{Asner:2008nq,Bevan:2014iga,Olsen:2015}. 
The quest for the pattern of $XYZ$ states can be addressed by finding 
partners to the presently observed states and measuring their radiative transitions. 
\panda can create any quantum number in $p\,\bar p$, 
which enables access to possible transitions between the family of $X$, $Y$ and
$Z$ states, and so map out their connections. 


It is a feature of many channels that there is an unexpected or unexplained enhancement at their threshold.
Some are likely direct reflections of QCD dynamics. It is therefore  important to be able to
perform dedicated threshold scans that unravel this dynamics. Similarly there are narrow states, for which
their width is unknown, presently limited by the resolution of the detectors. Scan experiments should be planned that determine 
their line shapes and width parameters and have the ability to distinguish the threshold behavior of different $J^{PC}$ values.
The high resolution provided by a cooled antiproton beam makes 
\panda the unique facility for such dedicated scans with the potential for further discoveries.


Moreover, in all sectors there is a rather limited knowledge of high-spin states, which may well be narrower and easier to identify. 
Such knowledge is an essential component of building an understanding of the nature of  the hadron spectrum in all its complexity.
The fascinating $XYZ$ states we presently see may very well be the glimpses of a new world of resonance physics,
whose study both theoretically and experimentally will lead to a deeper understanding of the strong dynamics of QCD.


\section*{Acknowledgments}
\noindent 
We thank the ExtreMe Matter Institute EMMI for significant financial support that made this event possible. 



\begin{thebibliography}{10}
\expandafter\ifx\csname url\endcsname\relax
  \def\url#1{\texttt{#1}}\fi
\expandafter\ifx\csname urlprefix\endcsname\relax\def\urlprefix{URL }\fi
\expandafter\ifx\csname href\endcsname\relax
  \def\href#1#2{#2} \def\path#1{#1}\fi

\bibitem{Edwards:2011jj}
R.~G. Edwards, {\em et~al.}, Phys. Rev. D84 (2011) 074508.


\bibitem{Brambilla:2004jw}
   N.~Brambilla, {\em et~al.}
   Rev.\ Mod.\ Phys.\  77 (2005) 1423.


\bibitem{Brambilla:2010cs}
   N.~Brambilla {\it et al.},
    Eur.\ Phys.\ J.\ C71 (2011) 1534.

\bibitem{GomezNicola:2001as}
A.~Gomez~Nicola, J.~R. Pelaez, Phys. Rev. D65 (2002) 054009.

\bibitem{Lutz:2003fm}
M.~F.~M. Lutz, E.~E. Kolomeitsev, Nucl. Phys. A730 (2004) 392.

\bibitem{Sanchis-Alepuz:2015hma}
H.~Sanchis-Alepuz, {\em et~al.}, Phys. Rev. D92 (2015) 034001.

\bibitem{Eichmann:2015zwa}
G.~Eichmann, {\em et~al.}, Acta Phys. Polon. Supp. 8 (2015) 425.

\bibitem{Fischer:2014xha}
C.~S. Fischer, {\em et~al.}, Eur. Phys. J. A50 (2014) 126.

\bibitem{Morningstar:1999rf}
C.~J. Morningstar, M.~J. Peardon, Phys. Rev. D60 (1999) 034509.

\bibitem{Dudek:2010wm}
J.~J. Dudek, {\em et~al.}, Phys. Rev. D82 (2010) 034508.

\bibitem{Adolph:2015pws}
C.~Adolph, {\em et~al.}, Phys. Rev. Lett. 115 (2015) 082001.

\bibitem{Kolomeitsev:2003ac}
E.~E. Kolomeitsev, M.~F.~M. Lutz, Phys. Lett. B582 (2004) 39.

\bibitem{Lang:2014yfa}
C.~B. Lang, {\em et~al.}, Phys. Rev. D90 (2014) 034510.

\bibitem{dsj_hybrids}
J.~Dudek, Phys. Rev. D84 (2011) 074023.

\bibitem{Lutz:2007sk}
M.~F.~M. Lutz, M.~Soyeur, Nucl. Phys. A813 (2008) 14.

\bibitem{Cleven:2014oka}
M.~Cleven, {\em et~al.}, Eur. Phys. J. A50 (2014) 149.

\bibitem{GarciaRecio:2003ks}
C.~Garcia-Recio, {\em et~al.}, Phys. Lett. B582 (2004) 49--54.

\bibitem{Kolomeitsev:2003kt}
E.~E. Kolomeitsev, M.~F.~M. Lutz, Phys. Lett. B585 (2004) 243.

\bibitem{Schat:2001xr}
C.~L. Schat, {\em et~al.}, Phys. Rev. Lett. 88 (2002) 102002.

\bibitem{Goity:2002pu}
J.~L. Goity, {\em et~al.}, Phys. Rev. D66 (2002) 114014.

\bibitem{Dashen:1993ac}
R.~F. Dashen, A.~V. Manohar, Phys. Lett. B315 (1993) 438.

\bibitem{Lutz:2010se}
M.~F.~M. Lutz, A.~Semke, Phys. Rev. D83 (2011) 034008.

\bibitem{Lutz:2014oxa}
M.~F.~M. Lutz, {\em et~al.}, Phys. Rev. D90 (2014) 054505.

\bibitem{Verduci:2014csa}
V.~Verduci, C.~B. Lang, {}\href {http://arxiv.org/abs/1412.0701}
  {\path{arXiv:1412.0701}}.

\bibitem{Crede:2013sze}
V.~Crede, W.~Roberts, Rept. Prog. Phys. 76 (2013) 076301.

\bibitem{Anisovich:2012}
A.~V. Anisovich, {\em et~al.}, EPJ A48 (2012) 15.

\bibitem{Sokhoyan:2015eja}
V.~Sokhoyan, {\em et~al.}, Phys. Lett. B746 (2015) 127.

\bibitem{Hartmann:2015kpa}
J.~Hartmann, {\em et~al.}, Phys. Lett. B748 (2015) 212.

\bibitem{Ablikim:2012}
M. Ablikim, {\em et~al.}, Phys. Rev. Lett. 110 (2012) 022001.

\bibitem{Aaij:2015tga}
R.~Aaij, {\em et~al.}, Phys. Rev. Lett. 115 (2015) 072001.

\bibitem{Lutz:2009ff}
{PANDA Collaboration}, {}\href {http://arxiv.org/abs/0903.3905}
  {\path{arXiv:0903.3905}}.

\bibitem{Brambilla:2014jmp}
N.~Brambilla, {\em et~al.}, Eur. Phys. J. C74 (2014) 2981.

\bibitem{Liu:2012jr}
L.~Liu, {\em et~al.}, Pos, (lattice 2012), 138.


\bibitem{Prelovsek:2015fra}
S.~Prelovsek, {}\href {http://arxiv.org/abs/1508.07322}
  {\path{arXiv:1508.07322}}.

\bibitem{Berwein:2015vca}
  M.~Berwein {\em et~al.}, {}\href {http://arxiv.org/abs/1510.04299}
  {\path{arXiv:1510.04299}}

\bibitem{hadrocharmonium}
M.~B. Voloshin, Phys. Rev. D87 (2013) 091501.

\bibitem{Wu:2013onz}
X.-G. Wu, {\em et~al.}, Phys. Rev. D89  (2014) 054038.


\bibitem{lange_charm13}
S.~Lange, {\em et~al.}, {}\href {http://arxiv.org/abs/1311.7597}
  {\path{arXiv:1311.7597}}.

\bibitem{prencipe_hadron15}
E.~Prencipe, {XVI International Conference on Hadron Spectroscopy}.

\bibitem{Braaten:2014qka}
   E.~Braaten, {\em et~al.},
   Phys.\ Rev.\ D90 (2014) 014044.

\bibitem{Guo:2014xrk}
F.-K. Guo, {\em et~al.}, {20th International Conference on Particles and Nuclei}.

\bibitem{b2kdd_babar}
B.~Aubert, {\em et~al.}, Phys. Rev. D77 (2008) 011102.

\bibitem{b2kdd_belle}
T.~Aushev, {\em et~al.}, Phys. Rev. D81 (2010) 031103.

\bibitem{Albaladejo:2015dsa}
M.~Albaladejo, {\em et~al.}, Eur. Phys. J. C75 (2015) 54.

\bibitem{pdg}
K.~A. Olive, {\em et~al.}, Chin. Phys. C38 (2014) 090001.

\bibitem{Polosa:2015} A.~Polosa, Phys. Lett. B746 (2015) 248.

\bibitem{Aaij:2014ala}
R.~Aaij, {\em et~al.}, Nucl. Phys. B886 (2014) 665.

\bibitem{Dudek:2014qha}
J.~J. Dudek, {\em et~al.}, Phys. Rev. Lett. 113 (2014) 182001.

\bibitem{Ryan:2015gda}
S.~M. Ryan, Lect. Notes Phys. 889 (2015) 35.

\bibitem{Sasaki:2015ifa}
K.~Sasaki, {\em et~al.},  Prog. Theor. Exp. Phys. (2015) 113B01.


\bibitem{Grube:2015fac}
B.~Grube, {\em et~al.}, {}\href {http://arxiv.org/abs/1510.07032}
  {\path{arXiv:1510.07032}}.

\bibitem{Adolph:2015tqa}
C.~Adolph, {\em et~al.}, {}\href {http://arxiv.org/abs/1509.00992}
  {\path{arXiv:1509.00992}}.

\bibitem{Kennedy-Spearman:1962}
J.~Kennedy, T.~D. Spearman, Phys. Rev. 126 (1962) 1596.

\bibitem{Eden_Landshoff_Olive_Polkinghorne}
R.~Eden, {\em et~al.}, Cambridge Univ (1966).

\bibitem{Lutz:2015lca}
M.~F.~M. Lutz, {\em et~al.}, Phys. Rev. D92 (2015) 016003.

\bibitem{Asner:2008nq}
D.~M. Asner, {\em et~al.}, Int. J. Mod. Phys. A24 (2009) S1--794.

\bibitem{Bevan:2014iga}
A.~J. Bevan, {\em et~al.}, Eur. Phys. J. C74 (2014) 3026.

\bibitem{Olsen:2015}
S. L. Olsen, {}\href {http://arxiv.org/abs/1510.07032}
  {\path{arXiv:1510.07032}}.
\end{thebibliography}

\end{document}